\documentclass[apj]{emulateapj}
\usepackage{amsmath}
\usepackage{natbib}
\usepackage{color}
\usepackage{url}

\newcommand{\clustername}{MACS~J1149+2223}
\newcommand{\snname}{``SN Refsdal''}
\newcommand{\clusterz}{0.543}
\newcommand{\zarcA}{1.491}
\newcommand{\zgal}{1.491}
\newcommand{\hst}{{\it HST}}

\newcommand{\J}{J14}
\newcommand{\implanerms}{$0\farcs15$}
\newcommand{\rmsone}{$0\farcs03$}

\newcommand{\lensgal}{G3}

\newcommand{\magA}{$18.5 ^{+6.4}_{-4.5}$}
\newcommand{\magB}{$14.4 ^{+7.5}_{-5.5}$}
\newcommand{\magC}{$20.5 ^{+19.1}_{-3.9}$}
\newcommand{\magD}{$11.0 ^{+6.9}_{-4.2}$}
\newcommand{\maggal}{$10.2 ^{+0.5}_{-1.5}$}

\newcommand{\timeB}{$  2 ^{+ 10}_{-  6}$}
\newcommand{\timeC}{$ -5 ^{+ 13}_{-  7}$}
\newcommand{\timeD}{$  7 ^{+ 16}_{-  3}$}
\newcommand{\timesnB}{$237 ^{+ 37}_{- 50}$}
\newcommand{\timesnC}{$-4251 ^{+369}_{-373}$}

\newcommand{\timeBls}{$  2 ^{+  6}_{-  4}$}
\newcommand{\timeCls}{$ -5 ^{+  7}_{-  4}$}
\newcommand{\timeDls}{$  7 ^{+  7}_{-  3}$}

\newcommand{\PxA}{$10.5 ^{+1.1}_{-1.4}$}
\newcommand{\PyA}{$-6.85 ^{+1.21}_{-0.83}$}
\newcommand{\PeA}{$0.61 ^{+0.01}_{-0.10}$}
\newcommand{\PthetaA}{$23.8 ^{+4.8}_{-2.3}$}
\newcommand{\PrcA}{$125 ^{+  7}_{- 15}$}
\newcommand{\PsigmaA}{$914 ^{+ 69}_{- 60}$}
\newcommand{\PxB}{$-13.2 ^{+1.6}_{-1.4}$}
\newcommand{\PyB}{$27.3 ^{+0.9}_{-4.4}$}
\newcommand{\PeB}{$0.30 ^{+0.05}_{-0.14}$}
\newcommand{\PthetaB}{$67.5 ^{+21.6}_{-5.5}$}

\newcommand{\PsigmaB}{$923 ^{+ 48}_{- 79}$}
\newcommand{\PeC}{$0.69 ^{+0.07}_{-0.10}$}
\newcommand{\PthetaC}{$33.6 ^{+6.0}_{-0.1}$}
\newcommand{\PrcC}{$9.59 ^{+0.41}_{-2.19}$}
\newcommand{\PsigmaC}{$416 ^{+ 17}_{- 28}$}
\newcommand{\PsigmaD}{$527 ^{+ 31}_{- 37}$}
\newcommand{\PxE}{$17.0 ^{+0.0}_{-0.5}$}
\newcommand{\PyE}{$101 \pm  1$}
\newcommand{\PthetaE}{$-54.4 ^{+1.0}_{-14.1}$}
\newcommand{\PsigmaE}{$355 ^{+ 60}_{-  2}$}
\newcommand{\PeF}{$0.36 ^{+0.40}_{-0.17}$}
\newcommand{\PthetaF}{$-34.2 ^{+2.9}_{-5.8}$}
\newcommand{\PrcF}{$0.16 ^{+2.93}_{-0.16}$}
\newcommand{\PsigmaF}{$235 ^{+ 87}_{- 11}$}
\newcommand{\PcutF}{$3.60 ^{+1.09}_{-1.58}$}
\newcommand{\PrcG}{$1.34 ^{+1.29}_{-1.31}$}
\newcommand{\PsigmaG}{$209 \pm 67$}
\newcommand{\PcutG}{$3.67 ^{+7.36}_{-1.92}$}

\newcommand{\nparameters}{38}
\newcommand{\nconstraints}{116}

\shorttitle{The lens model of \clustername}
\slugcomment{ApJ in prep: draft date \today}
\shortauthors{Sharon \& Johnson}

\begin{document}
\title{Revised Lens Model for the Multiply-Imaged Lensed
  Supernova, \snname~ in \clustername\altaffilmark{*}}
\altaffiltext{*}{Based on observations made with the NASA/ESA 
  {\it Hubble Space Telescope}, obtained from the Data Archive at the Space Telescope Science
  Institute, which is operated by the Association of Universities for
  Research in Astronomy, Inc., under NASA contract NAS 5-26555. These
  observations are associated with programs GO-9722, GO-12065 }
\author{Keren Sharon\altaffilmark{1}, 
{Traci L. Johnson\altaffilmark{1}},
} 

\email{kerens@umich.edu} 

\altaffiltext{1}{Department of Astronomy, University of Michigan, 1085 S. University Ave., Ann Arbor, MI 48109, USA}

\begin{abstract}
We present a revised lens model of \clustername, in which the first resolved
multiply-imaged lensed supernova was discovered. The lens model is
based on the model of Johnson et al. (2014) with some modifications. We
include more lensing constraints from the host galaxy of the newly
discovered supernova, and increase the flexibility of the model in
order to better reproduce the lensing signal in the vicinity of this
galaxy. The revised model accurately reconstructs the positions of the
lensed supernova, provides magnifications, and predicts the time delay
between the instances of the supernova. Finally, we reconstruct the
source image of the host galaxy, and position the supernova on one of its
spiral arms. Products of this lens model are available to the community
through MAST.

\end{abstract}


\section{Introduction}
A discovery of a multiply-imaged lensed supernova (SN)
has always been considered a low probability, high gain possibility in
SN surveys (e.g., Refsdal 1964, Kovner \& Paczynski 1988),
especially those focused on fields of lensing clusters. 
Although several lensed SN candidates have been reported (Quimby et al. 2014,
Patel et al. 2014,  Goobar et al. 2009), the recent 
discovery of a lensed SN in the field of
\clustername~ (Kelly et al. 2014a, 2014b) is the first reported case of
multiply-imaged one, whose instances are detected and resolved. 

\snname~ was discovered in \hst~ WFC3/IR data obtained as part of the Grism
Lens Amplified Survey from Space (GLASS; GO 13459, PI: Treu; { Schmidt et al. 2014}) between
2014 November 3 and 2014 November 20. We refer the reader to Kelly et
al. (2014b) for details of the discovery.
The host of \snname~ is a multiply-imaged face-on spiral galaxy at $z=$\zarcA,
lensed by the cluster halo into three full images and one partial
image (Smith et al. 2009), and further distorted by lensing
perturbations by cluster galaxies.  
As we will show in the following sections, \snname~ occurred in one of the spiral arms of this galaxy, which
happens to be going through secondary lensing by a cluster-member
galaxy. The SN is thus lensed into four images around that cluster galaxy,
in a configuration known as ``Einstein Cross'' (Kelly et al. 2014a). 

\clustername~ is an X-ray bright, strong lensing cluster at $z=$\clusterz~ (Ebeling et
al. 2007). Strong lensing models of \clustername~ were published by
Smith et al. (2009) and Zitrin et al. (2009) based on shallow \hst~
imaging (GO-9722, PI: Ebeling) and by Zitrin et al. (2011, 2014), and
Rau et al. (2014) based on data from the Cluster Lensing and Supernova
Survey with Hubble (CLASH) multicycle program
(PI: Postman; Postman et al. 2012). Smith et al. (2009) also measured
the spectroscopic redshifts of some of the lensed sources behind \clustername.
\clustername~ was recently selected as one of the Hubble Frontier
Fields (HFF, PI: Lotz) to be
deeply imaged by \hst. Preliminary lens models were derived by several teams
prior to the deep HFF imaging, using a close to uniform set of
constraints, but varying in lens modeling approaches, codes, and
assumptions (PI: Sharon -- Johnson et al. 2014; PIs: Kneib \&
Natarajan, CATS -- Richard et al. 2014;
PI: Brada{\v c}; PIs: Zitrin \& Merten; PI: Williams). 
The results of these models were made available to the 
community through Mikulski Archive for Space Telescopes
(MAST\footnote{\url{http://archive.stsci.edu/prepds/frontier/lensmodels/}}).  
While these models provide
a good description of the lens plane, they were not tailored to fit
any one lensed source in particular, but to provide the best mass
distribution and magnification estimates given a large set of
constraints coming from lensed 
galaxies at different redshifts. This approach allows a better
sampling of the slope of the mass distribution and facilitates studies of the
magnified background Universe. 

In this letter, we revise the model of Johnson et al. (2014; hereafter \J) and present
a new lens model for \clustername, that better
reproduces the lensing evidence at the vicinity of the multiply-imaged  
lensed SN.  
We assume a flat cosmology with $\Omega_{\Lambda}
= 0.7$, $\Omega_{m} =0.3$, and $H_0 = 70$  km s$^{-1}$
Mpc$^{-1}$. In this cosmology, $1\arcsec$ corresponds to 6.37 kpc at
the cluster redshift, $z=0.543$. Magnitudes are reported in the AB system.

\begin{figure}
\centering
\includegraphics[scale=0.38]{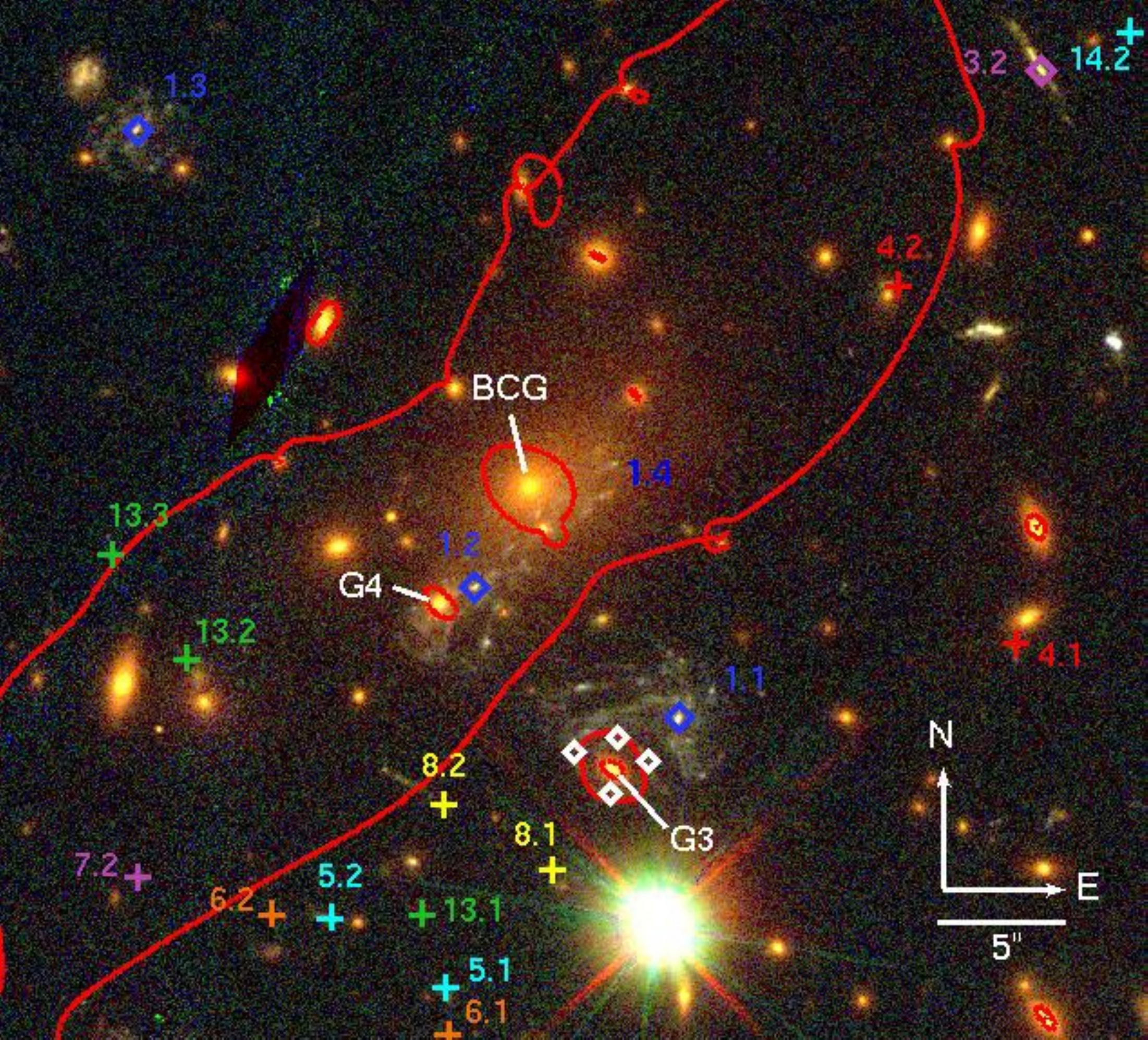}
\caption{{Color composite image of the core of strong lensing cluster
    \clustername~ from {\it HST}/ACS F814W (red), F606W (green),
    F435W (blue). The critical curves of the best-fit lens model are plotted in
red for $z=1.491$. 
The multiply imaged lensed sources that are used as constraints are
color coded and labeled, with diamonds for sources with spectroscopic
redshifts, and crosses for sources for which only photometric
redshifts are available. For a full list of these images, including
their coordinates, see \J.
The three full images and fourth partial image of
the host galaxy are labeled in blue, and the positions of the multiply-imaged
lensed SN are marked with white diamonds next to image 1.1. The
constraints within the images of source~1 are shown in 
Figure~\ref{fig.tile}}}
\label{fig.lensmodel}
\end{figure}

\begin{figure*}
\centering
\includegraphics[scale=0.3]{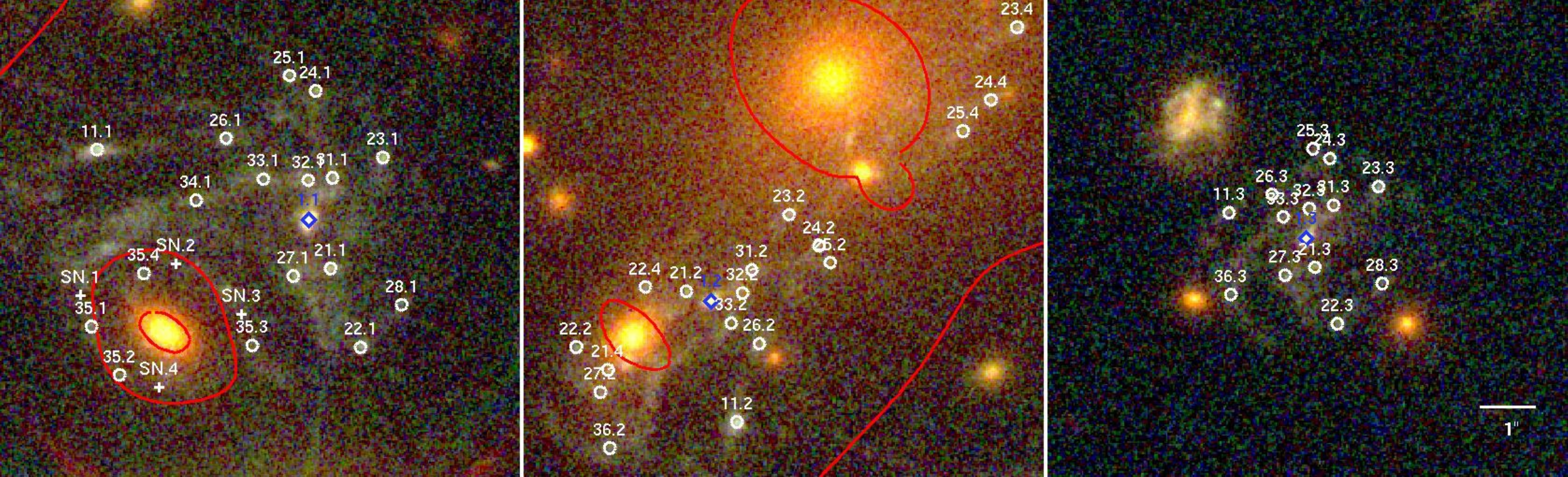}
\caption{{ Zoom in view on the three full images of the host galaxy;
    the emission knots that were used as constraints in this model are
    marked, as well as the SN positions from Kelly et al. (2014a).}}
\label{fig.tile}
\end{figure*}

\begin{figure*}
\centering
\includegraphics[scale=0.4]{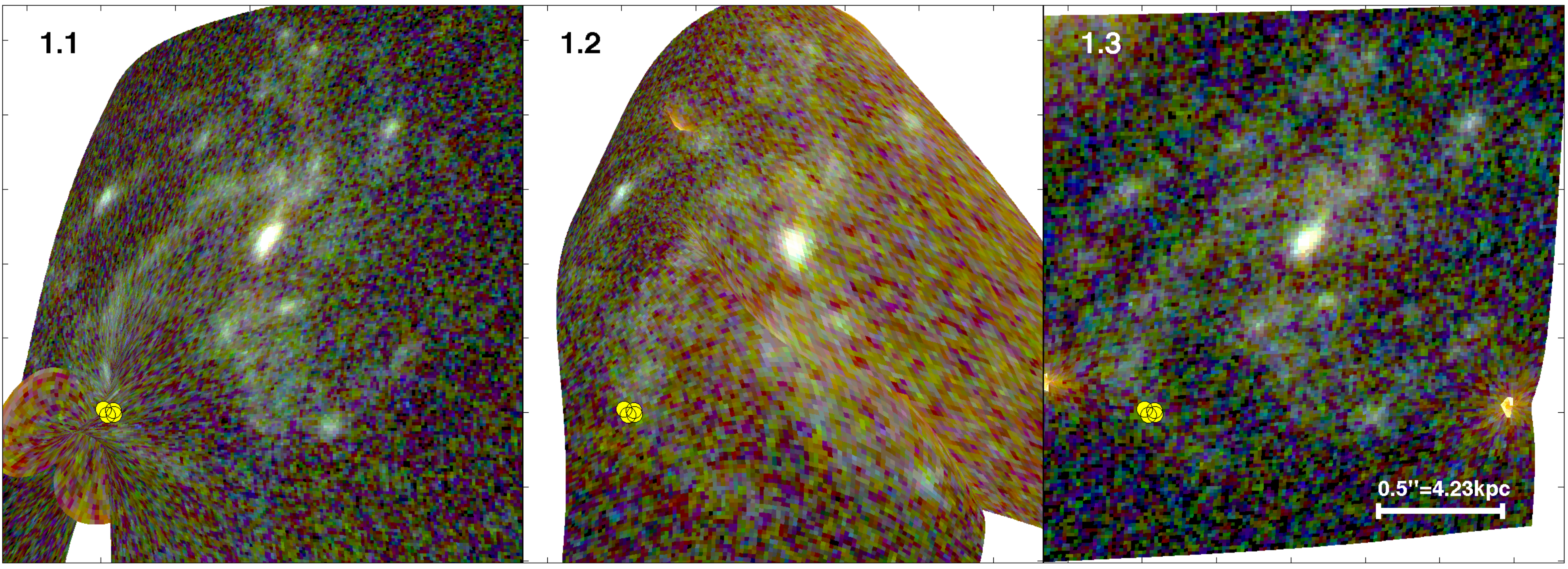}
\caption{{ Source reconstructions from the three full images of the
    host galaxy are in excellent agreement. The positions of
  the SN images are ray-traced from image 1.1 to the source plane, and marked in yellow
  circles on all the source images, indicating that it most likely
  occurred in one of the spiral arms. Ghost-like features appear as a result
of ray-tracing of light from the foreground galaxies.}}
\label{fig.source}
\end{figure*}

\begin{figure*}
\centering
\includegraphics[scale=0.4]{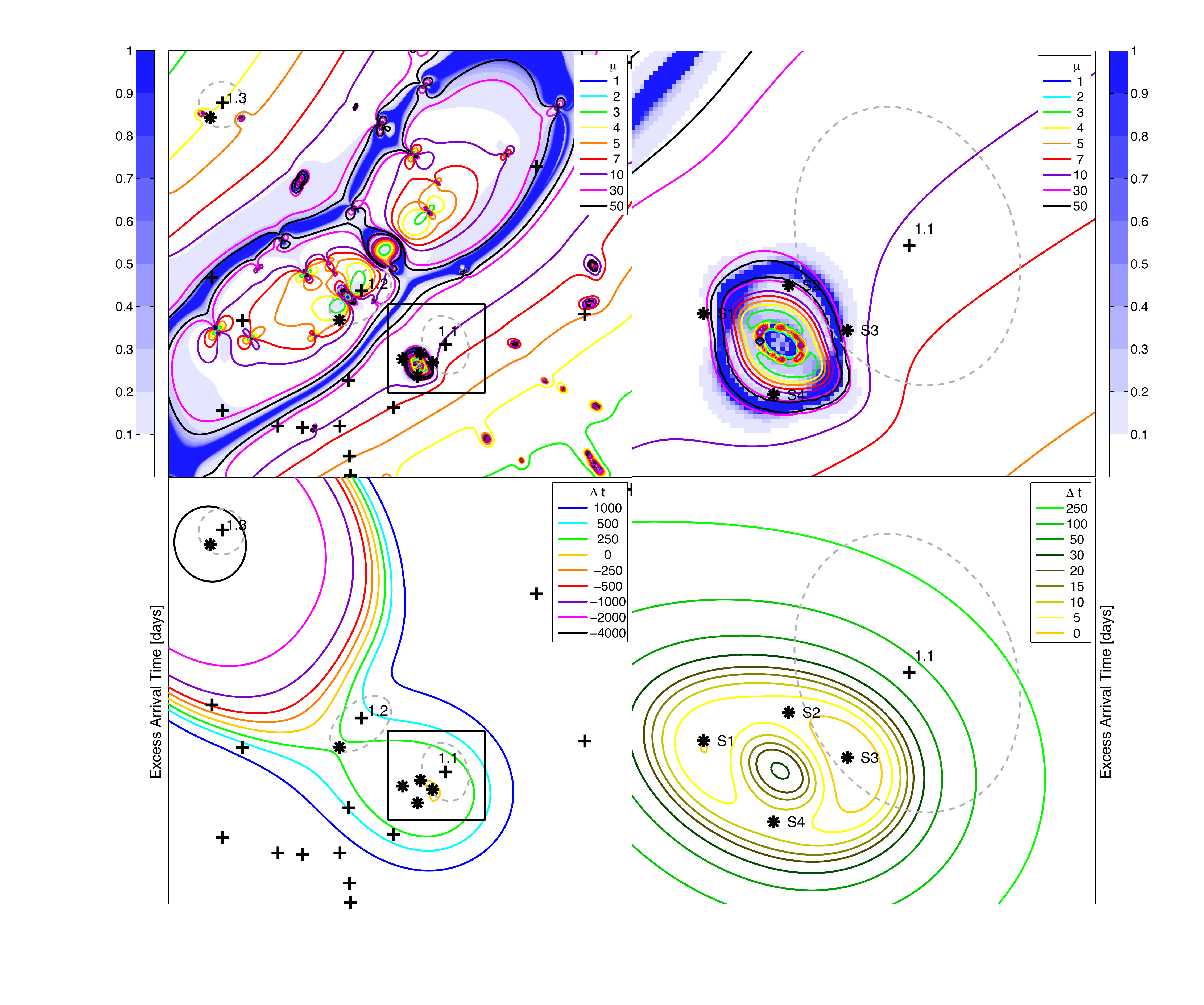}
\caption{{ The top panels show the magnification and its uncertainty for a source at
    $z=$\zgal.  
    The contours indicate the value of the 
    magnification, and the color gradient indicates the uncertainty in
    the magnification in each position. The uncertainty is given in
    units of $\Delta\mu/\mu$.  The field of view and scale are
    the same as those in Figure~\ref{fig.lensmodel} and
    Figure~\ref{fig.tile} for the left and right panels respectively.  To guide the
    eye, we plot the positions of the images of the lensed sources with
    black crosses, 
    and the SN positions in black asterisks. The predicted SN images
    are also marked. The general extent of the host galaxy is
    indicated in dashed ellipses. 
    The bottom panels show contours of the excess arrival time surface for
    light emitted from the source position in the source plane and
    traverses the lens plane. The excess arrival time is computed
    relative to S1, and given in days.}}
\label{fig.magnification}
\end{figure*}

\section{Lens Model}
We base our lensing analysis on the model presented by \J, with a few
modifications, as described below. The model is 
computed with Lenstool (Jullo et al. 2007), using Markov Chain Monte
Carlo (MCMC) sampling of the parameter space. 

The positional constraints of the arcs lensed by
\clustername~ are identical to those in \J~ (Figure~\ref{fig.lensmodel}), 
except for the main lensed source in
this cluster, the face-on spiral galaxy at $z=$\zgal~ that hosts
\snname. The
constraints that we used in \J~ did not favor this galaxy over other
lensed sources, and used as lensing evidence only the brightest emission
knot in each of the three full images of the source (1.1, 1.2 and 1.3
in Figures~\ref{fig.lensmodel} and \ref{fig.tile}).  Here,
we take full advantage of morphological 
information and use as constraints several emission knots on the
spiral arms of  each of
the lensed images of this galaxy. 
Figure~\ref{fig.tile} shows the lensing constraints that are used in
this model. 

The identification and mapping of the
knots between the lensed images of this galaxy are aided by the lens model of
\J, and constructed iteratively as higher levels of the current model  were computed.
Our identifications are done independently from those in
Smith et al. (2009) and Rau et al. (2014), but are in general agreement with their
identifications.
With our final model, many more features can be
identified (e.g., Rau et al. 2014); the features that are used as constraints here
provide a uniform spatial coverage of the galaxy.

To the morphological constraints we add
the positions of the four images of the SN from Kelly et al. (2014a), which appear around the
cluster galaxy \lensgal.

We note that the
bright emission knot next to G4 is not used as a constraint in this
model; the source reconstruction (see Section~\ref{s.results})
supports the association of this knot, as well as other clumps East
of G4,  with secondary lensing by
G4 that complicates image 1.2, as suggested by
Zitrin et al. (2009) and Rau et al. (2014). However, since this knot
is in close proximity to the critical curve of G4 we cannot determine
whether it is indeed a counter image of the core of the lensed galaxy,
or a different region that falls on the caustic and is therefore highly magnified. 
 In total, there are three full images of the host
galaxy, and additional images of some of the emission knots that form 
a partial forth image West of the
brightest cluster galaxy (BCG), and a partial image near G4. 

The mass distribution is composed of the five pseudo-isothermal
elliptical mass distribution (PIEMD) halos in \J, with
two additional halos that were individually optimized as described below.  
With the increased number of
constraints around the images of source~1, we increase the flexibility of
the model in these areas by freeing the parameters of the cluster galaxies closest
 in projection to 1.1 and 1.2. In particular, we allow the model to
fit for the ellipticity, position angle, core and cut radii, and velocity dispersion of the galaxy that
lenses the SN (\lensgal), as well as the core, cut radii and velocity
dispersion of the two central galaxies at the core of
the cluster (BCG, G4). We also free some parameters that were fixed in
\J~ to be fitted by the model; however, we find that these parameters
are consistent with the fixed values of \J. 
{ Generally, each halo represents a sum of the mass distribution of the 
galaxy and the dark matter halo in which it is embedded, and the lens model does
not distinguish between the two components. Thus, the best-fit parameters of some
galaxy scale halos (e.g., G1) may seem higher than typical for galaxies.}

Cluster galaxies are also included in the mass model as PIEMD halos,
with parameters scaled with their luminosity. 
\J~ used the cluster
galaxies from Smith et al. (2009), selected from $K$-band imaging data.
In order to increase the accuracy of the astrometric positioning of the
cluster galaxies, we construct our galaxy catalog directly from the
archival \hst~ imaging, taken as part of the CLASH multicycle program
(PI: Postman; Postman et al. 2012) and GO-9722 (PI: Ebeling; Smith et
al. 2009). 
Cluster galaxies are selected from the \hst~ ACS/$F606W$ and $F814W$
photometry as those with  $F606W-F814W$ colors that place 
them on the cluster red sequence in a color-magnitude diagram. We
assume $m_*=20.04$ in $ACS/F814W$. The coordinates, ellipticity, and
position angle of the cluster galaxies are fixed to their observed
values, and the other parameters are scaled with their
luminosity (see Limousin et al. 2005 for a description of the
scaling relations). 
Our final model includes \nconstraints~constraints and
\nparameters~free 
parameters (Table~\ref{table.lensmodel}). Among the
constraints are three spectroscopic redshifts of lensed galaxies,
measured by Smith et al. (2009). The
redshifts of the other nine lensed galaxies are set as free
parameters, with priors based on their photometric redshifts from Jouvel et al. (2013).

\section{Results and Conclusions}\label{s.results}
Figure~\ref{fig.lensmodel} shows the critical curves of the best fit
model over-plotted on a color image of the core of \clustername. 
We
find a significantly smaller scatter in the image plane for the
lensed galaxy that hosts \snname, with a typical scatter of  \implanerms~ for individual emission
knots. The image plane scatter of image 1.1 is reduced from $0\farcs83$ in \J~ to
\rmsone~ here. The lens model products are available to the community
through the HFF modeling page on MAST.
 
We reconstruct the source-plane image of the host galaxy using methods
described in Sharon et al. (2012, 2014), by ray-tracing the pixels of
each image of the lensed galaxy to the source plane.   
As can be seen in Figure~\ref{fig.source}, 
{ the source reconstruction of images~1.1 and 1.3 are in excellent
agreement. The north part of image~1.2 is also well matched to the
other two images. The south part of this image, in particular the
south spiral arm, is highly distorted in the lens plane by the cluster
galaxy G4. Thus, the reconstruction of the south part of image 1.2 is
not as well matched as it is in the other images. This may indicate
that the modeling of G4 can be improved upon to better constrain the
mass distribution of this galaxy.}

We ray-trace the position of the SN through the best-fit lens model
from the observed positions near \lensgal, to the other
images of the lensed galaxy. We find that the SN most likely occurred in one of the
spiral arms of this galaxy (Figure~\ref{fig.source}). 
The source reconstruction is in good agreement with the results of Rau
et al. (2014), who used a surface brightness distribution modeling
approach. 

Figure~\ref{fig.magnification} maps the magnification  ($\mu$)
and its fractional uncertainty  ($\Delta\mu/\mu$) for a source at $z=$\zgal. 
 We find
a magnification of $\mu(1.1)=$\maggal~ at the location of the brightest
emission knot of image 1.1; at the location of the four images of the
SN, the magnifications are somewhat higher, due to the added lensing
boost from \lensgal, $\mu(S1)=$\magA, $\mu(S2)=$\magB, $\mu(S3)=$\magC,
$\mu(S4)=$\magD, for images S1, S2, S3, S4,
respectively.  { Since the SN is practically a point source, microlensing by stars in the galaxy
\lensgal~ (Schechter \& Wambsganss 2002, Dobbler \& Keeton 2006),
as well as millilensing by substructure with mass of order of star clusters 
(Mao \& Schneider 1998, Metcalf \& Madau 2001, Dalal \& Kochanek 2002;
Nierenberg et al. 2014), 
may contribute to the magnification of some of the lensed images of the
SN. This could affect the flux ratios between
the SN images, and limit the use of flux ratios in modeling the
system. }

{ The arrival time surface for light emitted from the source
  location $\vec\beta$ and
  traverses a point $\vec\theta$ in the lens plane is given by (e.g., Schneider 1985)
\begin{equation}
\tau(\vec\theta,\vec\beta) = \frac{1+z_l}{c}\frac{D_{l}D_{s}}{D_{ls}}\bigg[\frac{1}{2}(\vec\theta-\vec\beta)^2-\psi(\vec\theta)\bigg],
\end{equation}
where 
$z_l$ is the lens redshift, $D_{l}$ and $D_{s}$ are the
  distances to the lens and the source, respectively, $D_{ls}$ is the
  distance from the lens to the source, and $\psi$ is the lensing potential.
 In the bottom panels of Figure~\ref{fig.magnification} we plot the
contours of the excess arrival time surface for light leaving the SN
source position in the source plane, measured relative to the
arrival time at the image-plane location of S1, $\Delta
t=\tau(\vec\theta,\vec\beta)-\tau(\textrm{S1}, \vec\beta)$.
Images occur at positions in the image plane
    in which the light travel time is stationary, i.e., minima, maxima and
    saddle points of this surface. Images S1, S3 and the predicted SN
    by 1.3 are minima, and images S2, S4 and the predicted SN by 1.2 are
    saddle points. }
The time delay predictions are given in days with respect
to image S1 of the SN:
{ $\Delta t_{\rm{S}2-\rm{S}1}=$\timeB, $\Delta t_{\rm{S}3-\rm{S}1}=$\timeC, $\Delta
t_{\rm{S}4-\rm{S}1}=$\timeD~ days.}
The time delay between S1 and the
predicted location of the SN in the other lensed images of the host galaxy  
are of order years, with $\Delta t_{1.2-\rm{S}1}=$\timesnB~days, 
and  $\Delta t_{1.3-\rm{S}1}=$\timesnC~days.
{The local maximum near the center of G3 indicates that 
    a demagnified SN image is predicted by the lens model, with time
    delay of $36^{+12}_{-8}$ days and $\mu=0.2\pm0.1$.
 The excess arrival time surface depends on the lensing potential
  and is highly sensitive to the inferred source plane
  position in this configuration. The source position is not an
  observed quantity, but is derived by the lens model in the
  modeling process. Each of the four images of the SN is ray-traced to
  a slightly different location in the source plane with a best-fit
  rms of $0\farcs02$.
To estimate the uncertainties on the time delay, 
we therefore consider both the model uncertainties, by
  computing the potentials from models drawn from the MCMC parameter
  sets as above, and
  the scatter in the source position of the lensed SN. To account for
  this scatter, we compute 15 arrival time surfaces for each set of
  MCMC parameters, each with a source position that is a weighted-mean
of the four computed SN source positions. The weights are set to either zero or
one for each SN, and all the combinations are considered, thus
sampling the area enclosed by the predicted source positions.

Our formal statistical uncertainties are drawn from a set of
parameters that represent the lowest $\chi^2$ steps in the chain that
approximately span $1\sigma$ in the parameter space. Since the
$\chi^2$ estimator marginalizes over all the lensing constraints, some
of these models result in a larger scatter in the source positions of
the SN, balanced by lower scatter in other constraints. If we consider
only the subset of models for which the SN source has a source plane
scatter of $<0\farcs05$, the uncertainties are reduced by 30-60\%: $\Delta
t_{\rm{S}2-\rm{S}1}=$\timeBls, $\Delta t_{\rm{S}3-\rm{S}1}=$\timeCls, $\Delta
t_{\rm{S}4-\rm{S}1}=$\timeDls~ days.

We find that the order of the arrival time is typically S3-S1-S2-S4,
with 69\% of the models predicting this order. 25\% of the models
predict arrival order of S1-S3-S2-S4. Most of the remaining models
predict that S3 appears first.
If considering only the models with low source-plane scatter, the
order S3-S1-S2-S4 is predicted by 93\% of the models. 
}

We note that the lens model does not use time delays as
constraints; at the time of this publication, time delays have not yet
been measured. 
The relatively short time delays that are predicted by the lens model
are consistent with the detection of three images of this SN at the
same time.

A small part of the
spiral arm in which the SN appears is lensed by the nearby 
cluster galaxy, \lensgal, forming a ring around it. The archival data
are not deep enough to uniquely identify substructure in this ring, but it will
likely be feasible with the full depth of the Hubble Frontier Field imaging. 
Nevertheless, the accurate positions of the newly discovered SN on
this ring were used as constraints to model the mass distribution in
this cluster galaxy; we find that the galaxy is well described by a
PIEMD halo centered on the observed light distribution
(Table~\ref{table.lensmodel}). 
We expect that the full depth of the HFF observation of \clustername~
will reveal ample new lensing constraints as new background sources
are detected. These constraints will be most useful for reducing the
uncertainties in the mass model of the cluster (e.g., Jauzac et
a. 2014a, 2014b) and of the galaxy G3. 
 
The discovery of a multiply imaged variable source with resolved
images has interesting implications on constraining the mass
distribution of its foreground lens. Follow up photometric observations of this
source will reveal the time delay between the images, and measure
 their relative magnifications.  
Adding these unique SN constraints 
to the positional constraints can improve
the lens model of \lensgal~ and the cluster in which it is embedded (e.g., Nordin et al. 2014). 
Such future analysis has the potential to constrain the truncation
radius of the galaxy due to tidal stripping, 
measure the slope of its mass distribution, and
the correlation between stellar mass and dark matter distributions, 
all of which can inform studies of galaxy evolution in cluster environment.

\begin{deluxetable*}{lccccccc} 
 \tablecolumns{8} 
\tablecaption{Best-fit lens model parameters  \label{table.lensmodel}} 
\tablehead{\colhead{Component }   & 
            \colhead{$\Delta$ RA ($\arcsec$)}     & 
            \colhead{$\Delta$ Dec ($\arcsec$)}    & 
            \colhead{$e$}    & 
            \colhead{$\theta$ (deg)}       & 
            \colhead{$r_{\rm core} $ (kpc)} &  
            \colhead{$r_{\rm cut}$ (kpc)}  &  
            \colhead{$\sigma_0$ (km s$^{-1}$)}             } 
\startdata 
cluster halo 1 (H1)         & \PxA       & \PyA       & \PeA  & \PthetaA        &\PrcA     & [1500]    & \PsigmaA  \\ 
cluster halo 2 (H2)         & \PxB       & \PyB        & \PeB   & \PthetaB          & [50] & [1500]     & \PsigmaB  \\ 
BCG (BCG)                       & [0]          & [0]           & \PeC& \PthetaC          & \PrcC & [200]     & \PsigmaC  \\ 
cluster galaxy 1 (G1)      & [25.6]       & [-32.2]        & [0.205] & [47.0]         & [0.233] & [40.2]     & \PsigmaD  \\ 
cluster galaxy 2 (G2)      & \PxE       & \PyE        & [0.8] & \PthetaE          & [0.261] & [300]     & \PsigmaE  \\ 
cluster galaxy 3 (G3)      & [3.162]       & [11.088]        & \PeF & \PthetaF          & \PrcF & \PcutF     & \PsigmaF  \\ 
cluster galaxy 4 (G4)      & [3.648]       & [-4.568]        & [0.355] & [48.3]          & \PrcG & \PcutG     & \PsigmaG  \\ 
L* galaxy  & \nodata & \nodata & \nodata & \nodata &  [0.15]  &     [30]&  [150]  \\

\enddata 
 \tablecomments{ All the coordinates are measured in arcseconds East and
   North of the center of the BCG, at [RA, Dec]=[177.39875
   22.398531]. Mass components are all PIEMD. The
   ellipticity of the projected mass density is expressed as $e=(a^2-b^2)/(a^2+b^2)$. $\theta$ is
   measured North of West. Error bars correspond to 1-$\sigma$
   confidence level as inferred from the MCMC optimization. Values in
   square brackets are for fixed parameters that were not optimized. The
   location and the ellipticity of the matter clumps associated with
   the cluster galaxies were kept fixed according to their
   light distribution, and the other parameters determined through
   scaling relations (see text). Halo and galaxy notations are adopted from \J. 
G3 is the cluster galaxy closest to image 1.1, and G4 is the cluster galaxy closest to
image 1.2.}
\end{deluxetable*}

\acknowledgments
This work makes use of the Matlab Astronomy Package (Ofek 2014). We
thank the annonymous referee for useful comments, which improved this manuscript.

\end{document}